\documentclass[preprint,aps,showpacs,amsmath,amssymb]{revtex4}
\usepackage{graphicx} %include figure files
\usepackage{dcolumn}%alingn table columns on decimal point
\usepackage{bm}%bold math

\begin{document}
\title{Spin transport of electrons through quantum wires with spatially-modulated
strength of the Rashba spin-orbit interaction}
\author{X. F. Wang}
\email{xuefeng@alcor.concordia.ca}
\affiliation{Concordia University Department of Physics\\
1455 de Maisonneuve Ouest \\
Montr\'{e}al, Qu\'{e}bec, H3G 1M8, Canada
}
\date{\today}

\begin{abstract}
We study  ballistic transport of spin-polarized electrons through quantum wires in which
the strength of the Rashba spin-orbit
interaction (SOI) is spatially modulated.
Subband mixing, due to SOI, between the two lowest
subbands is taken into account.
Simplified approximate expressions for the transmission are obtained
for electron energies close to the bottom of the first subband and near the value
for which  anticrossing of the two lowest subbands occurs.
In structures with periodically varied SOI strength,
{\it square-wave} modulation on the spin transmission is found when only
one subband is occupied and its possible application to the spin transistor is discussed.
When two subbands are occupied the transmission
is strongly affected by the existence
of SOI interfaces as well as by the subband mixing.
\end{abstract}

\maketitle

\clearpage

\section{INTRODUCTION}
With the development of nanotechnology, manipulation and measurement of
spin-orbit interaction (SOI) in semiconductor nanostructures
\cite{luo,hass,nit,eng,mat,hu1,row}
have been realized as well as injection and detection of spin-polarized current.
\cite{fied,jonk,ohn,poto,egu1,mire2,hu,lars,sch,ras2}
As a result, in the past years increasing attention has been drawn to
the spin-related behavior of quasi-one-dimensional (Q1D) electron
systems in the presence of SOI,
especially the Rashba SOI term,
of strength $\alpha$, which results from asymmetric electric confinement
in nanostructures. \cite{ras}
This has been also greatly stimulated by the proposal \cite{dat} of
establishing a spin transistor, among other novel spintronic devices, and its potential
application to the promising quantum computing. The SOI theory developed earlier for
bulk materials \cite{ras} and
two-dimensional electron systems \cite{byc} has been applied to
the electronic
band structure and spectral
properties of
realistic quantum wires. \cite{mor1,mor2,andr,gove1} Intriguing transport
properties through
quantum wires have been predicted as a result of peculiar features in their band structures
introduced by the SOI,
such as additional subband extrema and anticrossings. It has been
found that the spatial
distribution of the spin orientation in quantum wires can be greatly influenced by the
subband mixing and the existence of interfaces between different SOI strengths.
\cite{gove1}
Furthermore, the Rashba SOI has interesting effects on the
shot noise for spin-polarized and entangled
electrons, \cite{egue1}
and on the spectral properties of
interacting quantum wires. \cite{mor1} The former may lead
to another way of measuring the SOI strength in quantum wires.

To run a spin transistor based on the SOI
in a Q1D system,
a spin filter is required to provide the initial spin-polarized current. One of the
realistic options is to inject a spin current from
ferromagnetic
semiconductors or metals. \cite{fied,jonk,ohn,poto,egu1,mire2,hu,lars,sch,ras2}
The spin polarization of the injected
current is independent of the existence of
the SOI in the Q1D system \cite{mol} and this makes
it possible to separate the study of the
spin transistor from that of the spin filter.
Recently, several efforts have been made to describe in more detail the behavior of spin
polarized electrons in Q1D
systems in the presence of the Rashba SOI. A numerical
tight-binding simulation has been carried out to study ballistic transport \cite{mir}
through a quantum wire
in which one SOI segment is adiabatically connected to
two segments without
SOI.
The results illustrate that a uniform spin precession along the
wire should be observed provided the Rashba SOI strength is weak and
subband mixing is negligible. For strong SOI, however, spin modulation becomes
energy-dependent
and can be strongly suppressed at finite temperatures. Lately, a
square-wave modulation of spin polarization and a good spin transistor behavior
have been predicted in transport through periodically stubbed waveguides for
weak SOI and subband mixing due to it occuring 
only in the stubs.
\cite{wan}
Although in general subband mixing results in disagreeable effects on spin precession
in quantum wires,
it can offer further control of spin polarization at low temperatures in some situations.
\cite{wan,wan1,egue,mir}

In this paper we study ballistic
transport of Q1D spin-polarized electron gases in the presence
of a spatially-modulated Rashba SOI strength $\alpha$ and
take into account the subband mixing between the first and second subbands.
This modulation can be
achieved experimentally by external gates \cite{nit} and may result in further modulation
of spin currents,
as pointed out
in Refs. \onlinecite{mir} and \onlinecite{gove1}, and reported in Ref. \onlinecite{wan2} when subband mixing is neglected.
Different from
the periodically stubbed waveguides, with the same strength $\alpha$ everywhere,
studied previously, 
\cite{wan,wan1} here we consider waveguides without stubs but with
spatially-modulated $\alpha$, which, from an experimental point of view,
are easier to
realize and control.
We assume that the electric confinement, that gives rise to
the SOI, is strong enough that
excited states due to this confinement, as observed in Ref. \onlinecite{row}, 
are not occupied.

The paper is organized in the following way.
In Sec. II we propose a model  Hamiltonian with the Rashba SOI term to obtain
the band structure and wave function.
In Sec. III we formulate the transfer-matrix description of the transmission process
and in Sec. IV we present and discuss the results of spin transport.
Conclusions follow in Sec. V.

\section{MODEL}

\begin{figure}[tpb]
\includegraphics*[width=110mm,height=120mm]{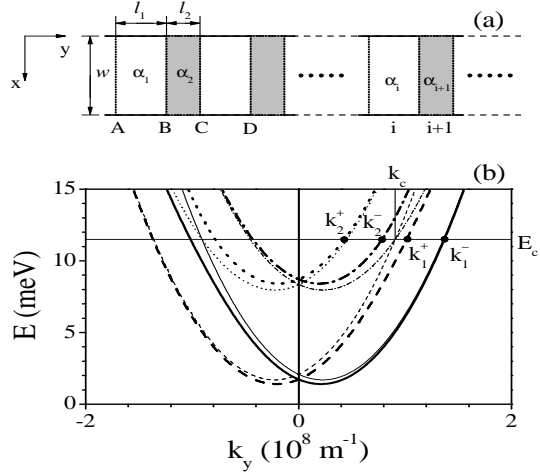}
\caption{(a) A quantum wire along the $y$ direction composed of a series of segments
$i=1,2,3,\cdots$
of SOI strength $\alpha_i$ and length $l_i$.
(b) Energy dispersion of the lowest two subbands in a
InGaAs quantum wire 600\AA \ wide. The thick (thin) solid, dashed, dotted,
dash-dotted curves present branches
$\varepsilon_1^-$ ($E_1^-$), $\varepsilon_1^+$ ($E_1^+$),
$\varepsilon_2^-$ ($E_2^-$), and $\varepsilon_2^+$ ($E_2^+$)
respectively.
}
\label{fig1}
\end{figure}

We consider a Q1D electron system, an
InGaAs/InAlAs quantum wire fabricated by confining a
two-dimensional (2D)
electron system in the $x$-$y$ plane by an infinitely high, square-well potential
$V(x)$. The wire has width $w$ along the $x$ direction, as shown in Fig. \ref{fig1}(a).
In the presence of the Rashba SOI the
Q1D one-electron Hamiltonian reads

\begin{equation}
\hat{H}=-\lambda {\bf \nabla}^{2}-i\alpha({\bf \sigma}\times {\bf \nabla})_{z}+V(x),
\label{hmlt}
\end{equation}
where $\lambda =\hbar ^{2}/2m^*$.
${\bf \nabla}=(\partial/\partial x,\partial/\partial y,0)$ 
is the Laplace operator, $\alpha $
is the strength of the SOI, and
${\bf \sigma}=(\sigma_{x},\sigma _{y},\sigma _{z})$
denotes the spin Pauli matrices.
In the $\sigma _{z}$ representation and with the use of the
eigenfunction of the Q1D Hamiltonian
$h(x)=-\lambda \nabla_{x}^{2}+V(x)$ without the Rashba term,
the eigenfunction is $\phi _{m}(x)=\sqrt{2/w}\sin (m\pi x/w)$ for $0\leq x\leq w$
and $m=1,2,3,\cdots$.
Then
the wave function of Eq. (\ref{hmlt}) can be expressed as 
$\Psi(k,{\bf r})=e^{iky}\sum_{m\sigma }\phi _{m}(x)
C_{m}^{\sigma}|\sigma\rangle$ with 
$|\sigma \rangle =(1,0)^T$ for spin up and
$(0,1)^T$ for spin down, with ${\cal X}^T$ denoting the transpose of
the column matrix ${\cal X}$.

For the sake of simplicity, while retaining the
subband mixing effects, we assume that only the
lowest two subbands ( $m=1$ and 2) are involved in the transport.
This can be considered as the actual case when the temperature and
the electron density are not too high. 
Then the secular equation $\hat{H} \Psi=E \Psi$
takes the form

\begin{equation}
\left[
\begin{array}{cccc}
E_{1}-E & \alpha k & 0 & -\delta \\
\alpha k & E_{1}-E & \delta & 0 \\
0 &  \delta & E_{2}-E & \alpha k \\
 -\delta & 0 & \alpha k & E_{2}-E
\end{array}
\right] \left(
\begin{array}{c}
C_{1}^{+} \\
C_{1}^{-} \\
C_{2}^{+} \\
C_{2}^{-}
\end{array}
\right) =0,
\label{hm}
\end{equation}
where $E_{m}=E_{m}^{0}+\lambda k^{2}$ and $E_m^0$ is
the $m$th subband bottom in the absence of SOI; 
$\delta =\alpha\int dx\phi_{2}(x)\phi _{1}^{\prime }(x)=8\alpha/3w$
is the mixing term due to SOI between
the first and the second subband. The resulting eigenvalues
$\varepsilon _{n}^{\sigma }(k)$ and eigenvectors $\Psi _{n}^{\sigma }(k)$ are
\begin{equation}
\left\{
\begin{array}{ll}
\varepsilon _{1}^{\pm}(k)=(E_1+E_2-\Delta E_{\mp})/2, & \ \ \ \ \
\ \ \ \Psi _1^{\pm}(k)=\frac{1}{F_{\pm k}} \left(
\begin{array}{c}
\phi _1\mp 2\delta \phi _2/B_{\pm k} \\
\pm\phi _1+ 2\delta \phi _2/B_{\pm k}
\end{array}
\right) , \\
\varepsilon _2^{\pm}(k)=(E_1+E_2+\Delta E_{\pm})/2, & \ \ \ \ \ \
\ \ \Psi _2^{\pm}(k)=\frac{1}{F_{\mp k}} \left(
\begin{array}{c}
\phi _2\mp 2\delta\phi_1/B_{\mp k} \\
\pm\phi _2+2\delta\phi _1/B_{\mp k}
\end{array}
\right).
\end{array}
\right.
\label{wf}
\end{equation}
Here $\Delta E_{\pm }=[(\Delta E_{12}\pm 2\alpha
k)^{2}+4\delta^2]^{1/2}$, $\Delta E_{12}=E_2^0-E_1^0$,
$B_{\pm k}=(\Delta E_{12}\mp 2\alpha k)+\Delta E_{-}$, and
$F_{\pm k}=[2+4\delta^2/B_{\pm k}^2]^{1/2}$.
Setting $\delta =0$ in Eq. (\ref{hm}) gives
the eigenfunctions without subband mixing
$|n,\pm\rangle=\phi_{n} (\pm 1, 1)^T/\sqrt{2}$ with corresponding energy
$E_n^{\pm}=E_{n}\pm \alpha k$. The resulting wavevector difference between the two spin
branches of each subband for
the same energy $E=E_n^+(k^+)=E_n^-(k^-)$,
which we denote as the intrasubband SOI splitting,
is constant for any energy and has the value
\begin{equation}
k_{\alpha }=k^--k^+=2m^*\alpha /\hbar^2.
\end{equation}

In Fig. \ref{fig1}(b) we plot the energy spectrum of the lowest two subbands with (thick
curves) and without (thin curves) the mixing term taken into account. This energy spectrum is essentially the same as that shown in
Fig. 2 of Refs. \onlinecite{wan1} and \onlinecite{egue}.
The intersections of the
Fermi energy, here taken equal to the anticrossing energy $E_{c}$
as shown by the horizontal bar,
with the energy spectrum define the wave vectors of the propagating modes.
$E_{c}$ is given by $E_{c}=E_{1}^{+}(k_{c})=
E_{2}^{-}(k_{c})=5E_{1}/2+9\hbar ^{2}E_{1}^{2}/(8m^*\alpha^{2})$
where the
branches without mixing $E_{1}^{+}$ and
$E_{2}^{-}$ ($E_{1}^{-}$ and
$E_{2}^{+}$) anticross
each other at the wave vectors
$k_{c}$ ($-k_{c}$). Without mixing, the
Fermi wave vector difference between the two spin branches of each subband
remains constant. With mixing, however, the difference between $\varepsilon _{n}^{-}$ and
$\varepsilon_{n}^{+}$ branches shows an energy dependence. Furthermore, in the branches
$E_{n}^{\pm }$ electrons have the same
spin orientation, but in the  branches $\varepsilon _{n}^{\pm }$,
which are not pure spin states,
the average electronic spins rotate continuously
from their low-energy orientation to their opposite high-energy orientation,
as also discussed recently in Ref. \onlinecite{gove1}.
At the anticrossing energy, the maximum mixing effect happens
between forward propagating electrons in
the $\varepsilon _{1}^{+}$
and $\varepsilon _{2}^{-}$ branches
(corresponding to $B_k=2\delta$ in Eq. (\ref{wf})) and between backward ones in
the $\varepsilon _{1}^{-}$ and $\varepsilon _{2}^{+}$ branches
($B_{-k}=2\delta$).
We denote the wave vector difference between the $\varepsilon _{1}^{\pm}$ and
$\varepsilon _{2}^{\mp}$ branches at this energy
as the intersubband SOI splitting
$2k_{\delta }$ related to $k_{c}$. For $\alpha < 2\hbar^2/m^* w$, we have
\begin{equation}
k_{\delta }\approx \frac{m^*\delta }{\hbar^2 k_{c}}=
\frac{8k_{\alpha}^2}{9\pi^2}w.
\end{equation}

\section{Transfer matrix}

We consider a quantum wire with a variable strength of SOI. It may be
composed of a series of SOI segments separated by SOI interfaces.
In each SOI segment, the SOI strength is approximately uniform and the Hamiltonian
described in the previous section applies.
To describe the electronic behavior propagating through the quantum wire,
we begin considering the transmission process of an electron with energy
$E$ through one SOI interface.
The electron is
incident from the left to the interface joining two segments (labelled $i$
and $i+1$) with different SOI strength $\alpha _{i}$ and $\alpha _{i+1}$
as shown in Fig. \ref{fig1}(a).
Taking into account only the lowest two
subbands, we write the wave function in segment $i$, in terms of
of the eigenfunction
$\Psi_{ni}^{\pm}$ given by Eq. (\ref{wf}) as
\begin{equation}
\varphi_{i}(x,y)=\sum_{\pm }\left[
c_{1i}^{\pm }\Psi _{1i}^{\pm }(k_{1i}^{\pm})e^{ik_{1i}^{\pm }y}
+\bar{c}_{1i}^{\pm }\Psi _{1i}^{\pm }(-k_{1i}^{\mp})e^{-ik_{1i}^{\mp }y}
+c_{2i}^{\pm }\Psi _{2i}^{\pm }(k_{2i}^{\pm})e^{ik_{2i}^{\pm }y}
+\bar{c}_{2i}^{\pm }\Psi _{2i}^{\pm }(-k_{2i}^{\mp})e^{-ik_{2i}^{\mp }y}\right].
\label{wf1}
\end{equation}
To obtain proceed we follow the approach of Ref.  \onlinecite{mol}:
we match  the wave function and its flux at
the interfaces between the $i$ and $i+1$ segments.
The velocity operator is given by
\begin{equation}
\hat{v}_{y}=\frac{\partial H}{\partial p_{y}}=\left[
\begin{array}{cc}
-i\frac{\hbar }{m^*}\frac{\partial }{\partial y} 
&\frac{\alpha }{\hbar } \\
\frac{\alpha }{\hbar } 
& -i\frac{\hbar }{m^*}\frac{\partial }{\partial y}
\end{array}
\right].
\end{equation}
The continuity of the wave function at the interface $y=y_{i,i+1}$, between
the $i$ and $i+1$ segments, gives $\varphi _{i}(x,y_{i,i+1})=\varphi
_{i+1}(x,y_{i,i+1})$ and that of the flux $\hat{v}_{y}\varphi
_{i}(x,y)|_{y_{i,i+1}}=\hat{v}_{y}\varphi _{i+1}(x,y)|_{y_{i,i+1}}$.
Multiplying the two equations by
$\Psi_{1i}^{* \pm }(k)$ and $\Psi _{2i}^{* \pm }(k)$ respectively, then
integrating over $x$\ we obtain eight linear equations for the eight
coefficients of the wave functions. Here we drop the subscript and superscript
pertaining to $k$ since they are the same as 
those for $\Psi$ as shown in Eq. (\ref{wf1}). Because electrons in branches
$\Psi_{1i}^{+}$ and $\Psi _{2i}^{-}$ are decoupled from electrons in branches
$\Psi _{1i}^{-}$ and $\Psi _{2i}^{+}$, a result of the symmetric property of
the wave function,
these equations are grouped into two similar but independent
equation groups with each composed of four equations.
The group corresponding to modes $\Psi _{1i}^{+}$ and
$\Psi_{2i}^{-}$ connects the column matrix 
$\hat{R}_{i}=(c_{1i}^{+},\bar{c}_{1i}^{+},c_{2i}^{-},\bar{c}_{2i}^{-})^T$
and column matrix 
$\hat{L}_{i+1}=(c_{1i+1}^{+},\bar{c}_{1i+1}^{+},c_{2i+1}^{-},\bar{c}_{2i+1}^{-})^T$
and reads:
\begin{equation}
\hat{S}_{i}\hat{R}_{i}=\hat{Q}_{i+1}\hat{L}_{i+1}.
\label{eqsq}
\end{equation}

Denoting the scalar product $\langle \Psi |\Psi ^{\prime }\rangle $ in all
matrix products by the integral $\int \Psi^{*T} (x,y)\Psi ^{\prime }(x,y)dx$ and
the direct product of the {\bf column} matrix
${\cal X}$ with the {\bf row} matrix ${\cal Y}$ by ${\cal X} \otimes {\cal Y}$, the upper ($2\times 4$) part of the
$4\times 4$ matrix $\hat{S}_{i}$ is given by 
${\cal A}_{i} \otimes {\cal B}_{i}$ and the lower part
 by
${\cal A}_{i}\otimes {\cal C}_{i}$
while the upper (2$\times $4) part of the 4$\times 4$ matrix $\hat{Q}_{i+1}$
is given by ${\cal A}_{i} \otimes {\cal B}_{i+1}$ and the lower part by
${\cal A}_{i}\otimes {\cal C}_{i+1}$. Here
${\cal A}_{i}=\left( \langle\Psi _{1i}^{+}(k)|,\langle \Psi _{2i}^{-}(k)|\right)^T $,
${\cal B}_{i}=\left( |\Psi _{1i}^{+}(k)\rangle ,|\Psi _{1i}^{+}(-k)\rangle ,|\Psi
_{2i}^{-}(k)\rangle ,|\Psi _{2i}^{-}(-k)\rangle \right) $ and
${\cal C}_{i}=\left( |\xi _{1i}^{+}(k)\rangle ,|\xi _{1i}^{+}(-k)\rangle ,|\xi
_{2i}^{-}(k)\rangle ,|\xi _{2i}^{-}(-k)\rangle \right) $ with 
$|\xi \rangle =\hat{v}_{y}|\Psi \rangle $.

If the SOI exists only in segment $i+1$\ ($\alpha _{i}=0$ and $\alpha
_{i+1}=\alpha $), we discard all the subscripts $i+1$ in the matrices $\hat{S}_{i}$
and $\hat{Q}_{i+1}$ and explicitly express them as 
\begin{equation}
\hat{S}_{i}=\left[ 
\begin{array}{cccc}
1 & 1 & 0 & 0 \\ 
0 & 0 & 1 & 1 \\ 
k_1^0/m^{\ast } & -k_1^0/m^{\ast } & 0 & 0 \\ 
0 & 0 & k_2^0/m^{\ast } & -k_2^0/m^{\ast }
\end{array}
\right]   \label{matrixs}
\end{equation}
and 
\begin{equation}
\hat{Q}_{i+1}=\sqrt{2}\left[ 
\begin{array}{cccc}
\frac{1}{\Theta _{1}} 
& \frac{1}{\bar{\Theta}_{1}} 
& \frac{2\delta }{\Theta_{2}\Lambda _{2}} 
& \frac{2\delta }{\bar{\Theta}_{2}\bar{\Lambda}_{2}} \\ 
-\frac{2\delta }{\Theta _{1}\Lambda _{1}} 
& -\frac{2\delta }{\bar{\Theta}_{1}\bar{\Lambda}_{1}} 
& \frac{1}{\Theta _{1}} 
& \frac{1}{\bar{\Theta}_{2}} \\ 
\frac{1}{\Theta _{1}}(\frac{\alpha }{\hbar^2 }+\frac{k_{1}^{+}}{m^{\ast }}) 
&\frac{1}{\bar{\Theta}_{1}}(\frac{\alpha }{\hbar^2 }-\frac{k_{1}^{-}}{m^{\ast }})
& \frac{2\delta }{\Theta _{2}\Lambda _{2}}(\frac{\alpha }{\hbar^2 }
+\frac{k_{2}^{-}}{m^{\ast }}) 
& \frac{2\delta }{\bar{\Theta}_{2}\bar{\Lambda}_{2}}(\frac{\alpha }{\hbar^2 }
-\frac{k_{2}^{+}}{m^{\ast }}) \\ 
\frac{2\delta }{\Theta _{1}\Lambda _{1}}(\frac{\alpha }{\hbar^2 }
-\frac{k_{1}^{+}}{m^{\ast }}) 
& \frac{2\delta }{\bar{\Theta}_{1}\bar{\Lambda}_{1}}(\frac{\alpha }{\hbar^2 }
+\frac{k_{1}^{-}}{m^{\ast }}) 
& \frac{1}{\Theta _{2}}(\frac{\alpha }{\hbar^2 }
-\frac{k_{2}^{-}}{m^{\ast }}) 
& \frac{1}{\bar{\Theta}_{2}}(\frac{\alpha }{\hbar^2 }
+\frac{k_{2}^{+}}{m^{\ast }})
\end{array}
\right] .  \label{matrixq}
\end{equation}
Here $k_1^0=%\sqrt{
[2m^{\ast }(E-E_{1}^{0})]^{1/2}/\hbar$ 
($k_2^0=%\sqrt{
[2m^{\ast}(E-E_{2}^{0})]^{1/2}/\hbar$) is the wave vector of the electrons
in the first (second)
subband in the segment without SOI and $\Lambda _{1}=B_{k_{1}^{+}}$, 
$\bar{\Lambda}_{1}=B_{-k_{1}^{+}}$, 
$\Lambda _{2}=B_{k_{2}^{-}}$, 
$\bar{\Lambda}_{2}=B_{-k_{2}^{-}}$, 
$\Theta _{1}=F_{k_{1}^{+}}$, 
$\bar{\Theta}_{1}=F_{-k_{1}^{+}}$, 
$\Theta _{2}=F_{k_{2}^{-}}$, and 
$\bar{\Theta}_{2}=F_{-k_{2}^{-}}$ are defined in the segment
with SOI of strength $\alpha$.

The above complex matrix can be simplified approximately in the low-energy
limit ($E\gtrsim E_1^0$) and the anticrossing energy limit
($E\thickapprox E_c$). If the electron density is sufficiently low that the Fermi
energy is close to the bottom of the first subband $E_{1}^{0}$ in
segment $i$, the correction to the wave function caused by subband mixing is
negligible or we have $\Theta=\bar{\Theta}=\sqrt{2}$ and $\delta =0$ in Eqs. 
(\ref{matrixs}) and (\ref{matrixq}). We find all the spin modes in the quantum
wire are decoupled from each other and the transfer equations for all modes
have similar forms. The transfer equation for the mode $\Psi_{1}^{+}$ has
the form 
\begin{equation}
\left[ 
\begin{array}{cc}
1 & 1 \\ 
k_1^0 & -k_1^0
\end{array}
\right] \left( 
\begin{array}{c}
c_{1i}^{+} \\ 
\bar{c}_{1i}^{+}
\end{array}
\right) =\left[ 
\begin{array}{cc}
1 & 1 \\ 
K_{2} & -K_{2}
\end{array}
\right] \left( 
\begin{array}{c}
c_{1i+1}^{+} \\ 
\bar{c}_{1i+1}^{+}
\end{array}
\right),  \label{le}
\end{equation}
with $K_{2}=(1/\hbar )%\sqrt{
[2m^*(E-E_{1}^{0}+\varepsilon
_{0}+V_0)+(m^*\alpha /\hbar )^{2}]^{1/2}$ and $\varepsilon _{0}\approx \delta
^{2}/\Delta E_{12}$ the energy correction to the first subband as a
result of the SOI subband mixing. $V_0$ denotes the conduction band offset
in the segment $i+1$ reckoned from the conduction band bottom in segment $i
$, which may be introduced by material mismatch at the interface or by an
external gate bias.

When the SOI strength is in the range $\hbar ^{2}/m^{\ast }w<\alpha <2\hbar
^{2}/m^{\ast }w$, the anticrossing energy is higher then the second subband
bottom but lower than the third subband bottom and the
intersubband SOI splitting is much smaller than the intrasubband SOI splitting,
i.e. $k_{\delta }\ll k_{\alpha }$. Eqs. (\ref{matrixs}) and (\ref{matrixq})
can be greatly simplified if the electron energy is near to the anticrossing
energy, i.e. $E\simeq E_{c}$.
In this case, we use the approximation  $V_0=0$,
$\Lambda _{1}\approx 2\delta \sqrt{k_{2}^{-}/k_{1}^{+}}$, 
$\Lambda _{2}\approx 2\delta \sqrt{k_{1}^{+}/k_{2}^{-}}$, 
$\bar{\Lambda}_{1}\approx 8\alpha k_1^0$, 
$\bar{\Lambda}_{2}\approx 8\alpha k_2^0$, 
$\Theta _{1}\approx 2\sqrt{k_{1}^{+}/k_1^0}$, 
$\Theta _{2}\approx 2\sqrt{k_{2}^{+}/k_2^0}$,
$\bar{\Theta}_{1}\approx \bar{\Theta}_{2}\approx \sqrt{2}$, 
$k_{1}^{+}\approx k_{c}+k_{\delta}$,
$k_{2}^{-}\approx k_{c}-k_{\delta}$,
$k_{1}^{-}\approx k_{c}+k_{\alpha }$, 
$k_{2}^{+}\approx k_{c}-k_{\alpha }$, 
$k_1^0\approx k_{c}+k_{\alpha }/2$,
$k_2^0\approx k_{c}-k_{\alpha }/2$,
$\delta /\bar{\Lambda}\approx 0$. Eq. (\ref{eqsq}) reduces to 
\begin{equation}
\left( 
\begin{array}{c}
c_{1i}^{+} \\ 
\bar{c}_{1i}^{+} \\ 
c_{2i}^{-} \\
\bar{c}_{2i}^{-}
\end{array}
\right) =\left[ 
\begin{array}{cccc}
\sqrt{k_{1}^{+}/2k_1^0} & 0 & \sqrt{k_{2}^{-}/2k_1^0} & 0 \\ 
0 & 1 & 0 & 0 \\
-\sqrt{k_{1}^{+}/2k_2^0} & 0 & \sqrt{k_{2}^{-}/2k_2^0} & 0 \\ 
0 & 0 & 0 & 1
\end{array}
\right] \left( 
\begin{array}{c}
c_{1i+1}^{+} \\ 
\bar{c}_{1i+1}^{+} \\ 
c_{2i+1}^{-} \\ 
\bar{c}_{2i+1}^{-}
\end{array}
\right)
\label{cross} .
\end{equation}

We see that, in the anticrossing energy limit, the mixing happens mainly
between the two modes involved in the anticrossing ($c^+_1$ and $c^-_2$) and the modes
corresponding to the coefficients $\bar{c}_{1}^{+}$ and  $\bar{c}_{2}^{-}$
remain almost intact when
transmitting from segment $i$\ to segment $i+1$ though there is a
wave vector mismatch between the two segments.

Once the matrices $\hat{S}_{i}^{-1}$ and $\hat{Q}_{i+1}$ in Eq. (\ref{eqsq}) are known,
the transfer matrix for the interface
joining segment $i$ and $i+1$ is obtained simply: $M_{i,i+1}=\hat{S}_{i}^{-1}\hat{Q}_{i+1}$.
For a quantum wire with $n$ segments, the total transfer matrix then reads
\begin{equation}
\hat{M}=\hat{P}_{1}\prod_{i=1,n-1}\hat{M}_{i,i+1}\hat{P}_{i+1},
\label{tmatrix}
\end{equation}
where the transfer matrix for the $i$-th segment of length $l$\ is expressed as
\begin{equation}
\hat{P}_{i}=\left[
\begin{array}{cccc}
e^{-ik_{1i}^{+}l} & 0 & 0 & 0 \\
0& e^{ik_{1i}^{-}l} & 0 & 0 \\
0& 0 & e^{ik_{2i}^{-}l} & 0 \\
0& 0 & 0 & e^{-ik_{2i}^{+}l}
\end{array}
\right] .
\end{equation}

A transfer matrix similar to $\hat{M}$ given by Eq. (\ref{tmatrix}) is obtained for the modes
$\Psi_{1i}^{-}$ and $\Psi _{2i}^{+}$ by applying the same process as above and will not be
shown here. Finally a $8\times 8$ transfer matrix $\hat{T}$ is obtained
connecting the wave-function coefficients of the electron with energy $E$ at the
left inlet
end, 
$\hat{I}=(c_{1I}^{\pm },\bar{c}_{1I}^{\pm },c_{2I}^{\pm },\bar{c}_{2I}^{\pm })^T$,
and at the
right outlet
end, 
$\hat{O}=(c_{1O}^{\pm },\bar{c}_{1O}^{\pm },c_{2O}^{\pm },\bar{c}_{2O}^{\pm })^T$ in the form
$\hat{I}=\hat{T}\hat{O}$.

\section{results and discussion}

In the following we present results for ballistic transport
of electrons, incident with spin up (along
the $z$ direction),
through a quantum wire in which segments with and without SOI
alternate periodically.
If not otherwise specified, zero temperature and parameters $w=600$ \AA,
$l=2500$ \AA, $\alpha =3.45\times 10^{-11}$ eVm, $m^{*}=0.05$, and $V_{0}=0$
will be assumed.
The strength of SOI can be adjusted experimentally \cite{nit,eng,mat,hu1}
and here we use the same value of $\alpha$ as used in Ref. \onlinecite{egue}
for the sake of comparison.
At zero temperature, only electrons of the Fermi energy
contribute to the transport. In the inlet and outlet
segments of the quantum wire, we assume there is no
SOI and the group velocity of the electrons
is proportional
to the wave vector 
($k_n^0=[2m^*(E-E_{n}^{0})]^{1/2}/\hbar,\, n=1,2$) and the
density of states to its inverse. When using the
normalized coefficients for the input wave function
$(c_{1I}^{\pm }=1/\sqrt{2k_1^0},c_{2I}^{\pm }=1/\sqrt{2k_2^0})$,
we express the zero-temperature spin-up
(spin-down) partial conductance $G_{n}^{+}(G_{n}^{-})$ and the partial transmission 
$T_{n}^{+}(T_{n}^{-})$ at the outlet end via the subband $n$ and the reflection
$R_n^+$ ($R_n^-$) from the inlet end as:
\begin{equation}
G_{n}^{\pm }=\frac{e^{2}}{h}T_{n}^{\pm }
=\frac{e^{2}}{h}\frac{k_{n}}{2}\left| c_{nO}^{+}\pm
c_{nO}^{-}\right| ^{2}; \, \,\,\,\,\, R_n^{\pm}
=\frac{k_{n}}{2}\left| \bar{c}_{nI}^{+}\pm
\bar{c}_{nI}^{-}\right| ^{2}.
\end{equation}
The finite-temperature conductance $G^{\pm}_n(T)$ is obtained by
integrating over energy the above zero-temperature
conductance multiplied by the Fermi distribution function $f_n$:
\begin{equation}
G_n^{\pm}(T)=-\int dE G_n^{\pm}(E,T=0)\frac{df_n(E,T)}{dE}.
\end{equation}

\subsection{The low-energy limit}

If only the first subband is occupied and $E_c$ and $E_2^0$
are much above the Fermi energy,
Eq. (\ref{le}) can be used to estimate the transmission through the quantum wire.
In a quantum wire with only one SOI segment, of strength $\alpha _{2}=\alpha$ and
length $l_{2}=l$, sandwiched between two segments without SOI as the part A-B-C-D shown in Fig. \ref{fig1}(a),
the total transmission takes the form
\begin{equation}
T=\frac{t}{t\cos ^{2}(K_{2}l)+\sin ^{2}(K_{2}l)}
\label{lwegt}
\end{equation}
where $t=4(k_1^0)^2K_{2}^{2}/[(k_1^0)^2+K_{2}^{2}]^{2}$. This is a
sinusoidal dependence with a maximum $T_{max}=1$ for
$\sin (K_{2}l)=0$ and a minimum $T_{min}=t$. For a quantum wire of fixed width,
the stronger is the SOI strength and the less is the energy of the incident
electrons, the more efficient is the modulation of the transmission.
Another feature of this 
interface-induced transmission modulation is that
it does not affect the spin polarization. The output percentage of spin-up and
spin-down electrons remains the same as predicted neglecting the interface
effect, $T^{+}=T\cos ^{2}(\theta /2)$ and $T^{-}=T\sin ^{2}(\theta /2)$
with $\theta =k_{\alpha }l$. These features offer the possibility of independent control
of the total transmission and of the spin polarization and will be advantageous when designing
a spin transistor employing this system.

\begin{figure}[h]
\includegraphics*[width=100mm,height=120mm]{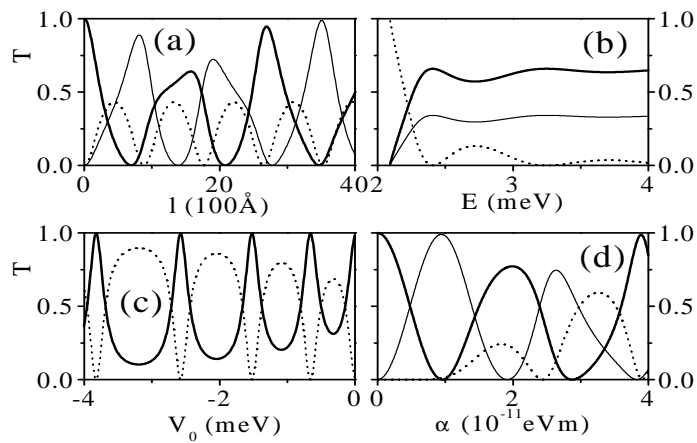}
\caption{Transmission $T^+$ (thick solid curves) and $T^-$ (thin solid curves),
and reflection $R^+$ (dotted curves) of a quantum wire with one SOI segment
as a function of 
(a) $l$,
(b) $E$,
(c) $V_0$ of the SOI segment of length $l=2800$\AA, and 
(d) $\alpha$ when $E=\varepsilon_1(0)+0.1$meV.
The result in (a) and (c) corresponds to the electrons
of energy $E=\varepsilon_1(0)+0.2$meV with $\varepsilon_1(0)=2.09$meV.
}
\label{fig2}
\end{figure}

In Fig. \ref{fig2}, we plot $T^{+}$ (thick solid curves), $T^{-}$
(thin solid), and $R^{+}$ (dotted) as a function of the main
parameters of the quantum wire.
Since only spin-up electrons are incident and there is no spin-flip mechanism,
spin-down reflection
is not observed. The
spin-up and spin-down electron output flux show, in Fig. \ref{fig2}(a),
a modulated sinusoidal
dependence on the length $l$ of the SOI segment instead of the simple one
as the interface effect is neglected. \cite{dat} For electrons near
the bottom of the first subband,
a strongly energy-dependent reflection happens (Fig. \ref{fig1}(b)) due to the wave vector
mismatch between electrons in segments with and without SOI.
A similar energy dependence of the conductance (transmission)
has also been found in Ref. \onlinecite{mir}.
In some cases, e.g., when an extra gate bias is applied,
an offset $V_0$ exists between the conduction bands
of the material in the segment with SOI and the one
without SOI and can be adjusted.
This may introduce further modulation to
the spin transmission. 
Fig. \ref{fig2}(c) illustrates the oscillatory dependence of the
transmission and
the reflection on this offset.
An increasing amplitude of the oscillation is found for a lower conduction-band
bottom of the SOI segment.
Here $l=2800$\AA\ is chosen so that no spin-down transmission is observed.
An oscillatory dependence of the transmission
and reflection is also found as a function of the SOI strength as shown
in Fig. \ref{fig2}(d).
For $\alpha > 1.5\times 10^{-11}eVm$ the reflection
of low energy electrons may become significant.

\begin{figure}[h]
\includegraphics*[width=110mm,height=110mm]{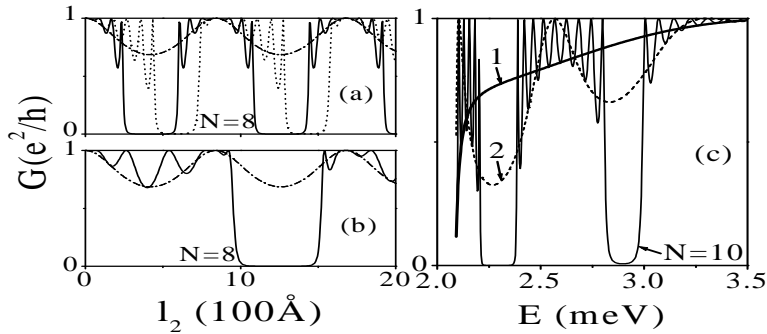}
\caption{(a) Total conductance of a periodic structure of 8 identical
units as functions of the length $l_2$ of the SOI segment when
fixing the length $l_1$ of the segment without SOI at $l_1=2375$\AA\  (solid
curve) and $l_1=2000$\AA\  (dotted curve). (b) The same as
(a) when the total length of each unit is fixed to
$l_1+l_2=2100$\AA. The dash-dotted curves in (a) and (b) are results for one unit cases.
(c) Conductance in a periodic structure of $N$ identical
units as functions of the electron energy when $l_1=1350$\AA\  and $l_2=1100$\AA.
The numbers beside the curves label $N$.
} \label{fig3}
\end{figure}

Now we introduce a periodic structure consisting of identical units that are repeated
along the wire. Each unit is composed of one non-SOI segment of
length $l_{1}$\ and one SOI segment of strength $\alpha $ and length $l_{2}$
as shown in Fig. \ref{fig1}(a).
Electrons are incident at the left end and exiting at the other; both ends
are segments without SOI. Because the Fabry-Perot-like interference of electron 
waves
happens between interfaces connecting regions with different SOI strengths and then
different energy-momentum dispersion relations, the transmission minima of one SOI segment described by Eq. (\ref{lwegt})
deepen with the increase of the number of units and
transform into transmission gaps when the number is big enough. This happens in a similar
way as in the Krong-Penney model of solids. As shown in Fig. \ref{fig3}(a), where each curve corresponds to a fixed $l_1$, almost square wave curves (solid and dotted curves) as
functions of $l_2$ are observed in structures of 8 units, comparing with the sinusoidal
total transmission $T$ (dash-dotted curves) for one unit. For structures with more units,
increased frequency and amplitude of the oscillations between the gaps are observed.
The position and width of the conductance gaps shift when varying $l_1$.
If we fix the total
length $l_{1}+l_{2}$ of each unit, similar square wave transmission is obtained as
shown in fig. 3(b) but with different gap width.
The percentage of the spin-up and spin-down conductance
here depends only on the total length of the SOI segment and
$k_{\alpha}$ and can be easily figured out in the
same way as that for one-unit quantum wires.

%fig.3

A similar square-wave conductance can also be found as a function of the electron energy.  For a quantum wire with only one SOI segment, the transmission increases
monotonically as a function of the energy and approaches unity when the electron wave vector is much larger than $k_{\alpha}$ as shown by the thick curve in Fig. \ref{fig3}(c).
A simple oscillating conductance (the dotted curve) appears in a quantum
wire with two SOI segments; 
square conductance gap can be observed in a wire with 10 (thin solid curve) or more units
and miniband develops in a SOI superlattice as appears generally in periodic structures.

\subsection{Two-subband transmission}

Electrons can propagate via the second subband when their energy is high enough.
In the trivial case of weak SOI strength that subband mixing is negligible, each mode propagates through the quantum
wire almost independently, the spin transmission of electrons in the second subband can be estimated in a similar
way as in the first subband by Eq. (\ref{le}). Results similar to
those obtained in the low-energy
limit of the first subband are also obtained for the second subband and will not be shown here in detail.
In the following
we will concentrate on the SOI-interface effect on the two-subband transmission when the subband mixing is important.

%fig4
\begin{figure}[h]
\includegraphics*[width=70mm,height=90mm]{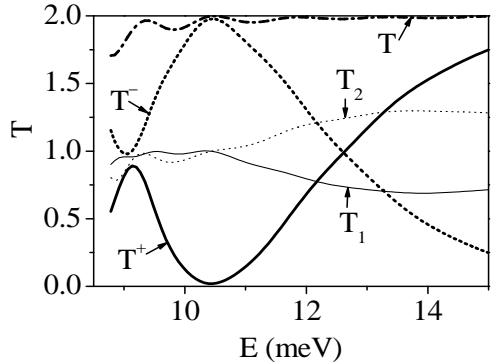}
\caption{Transmission vs electron energy when two
subbands are occupied in a quantum wire with one SOI segment of
length $l=2000$\AA. The total transmission $T$ (dash-dotted curve) and partial
transmissions corresponding to different
spin orientations $T^{\pm}$ (thick solid for $+$ and thick dotted for $-$)
and different subbands $T_n$ (thin solid and dotted for $n=1$ and $2$ respectively)
are shown separately.
The bottom of the second subband in the SOI segment is about 8.8meV.
}
\label{fig4}
\end{figure}

In Fig. \ref{fig4} we show the energy dependence of the total
transmission $T$ (dash-dotted curve), spin transmission
$T^{\pm}$ (thick solid and dotted curves) and transmission
from the first $T_1=T_1^++T_1^-$
and the second $T_2=T_2^++T_2^-$ subband (thin curves).
The reflection becomes significant mainly for those electrons
propagating via the second subband
when the electron energy is close to the second subband bottom,
$\varepsilon_2(0)$, and is not as strong as observed
for electrons of energy near the first-subband bottom.
At higher values of energy, the reflection become negligible and
more electrons come out from the second-subband. As a result of
the subband mixing and corresponding energy band
modification, the percentage of spin-up and spin-down electrons
is strongly dependent on the electron energy as also reported
in Ref. \onlinecite{mir}. 
In quantum wires with multiple SOI segments, the transmission
is further modulated for the same reason as discussed in the
one subband case but the modulation is much more complex and
irregular as a result of the coupling between different modes at interfaces. 

In a Q1D electron system formed from an ideal 2D system, where the SOI strength is
independent of the electron density as studied in this paper,
the carrier density dependence of the ballistic conductance (transmission) can easily
be figured out from their energy dependence plotted in Fig. \ref{fig4}. However,
in a realistic semiconductor system \cite{nit} the SOI strength and the quantum wire width
may vary as the carrier density changes.
The density dependence of the conductance should be estimated
using a more realistic model taking these effects into account.
Another point
concerning realistic systems
is the effect of excited states due to the confinement along $z$ direction, which introduces
the SOI and forms
a quasi 2D system rather than an ideal 2D system in the $x$-$y$ plane. As observed in
Ref. \onlinecite{row}, the excited states of the quasi 2D system can be occupied when the carrier
density is high. In a simple
approximation, this case can be treated as a two-carrier system where the electrons in the ground state and the excited state transport independently if the confinement along $x$ direction is symmetric. As a result, the conductance will be enhanced because more channels are opened to transport electrons.

%fig5
\begin{figure}[h]
\includegraphics*[width=70mm,height=90mm]{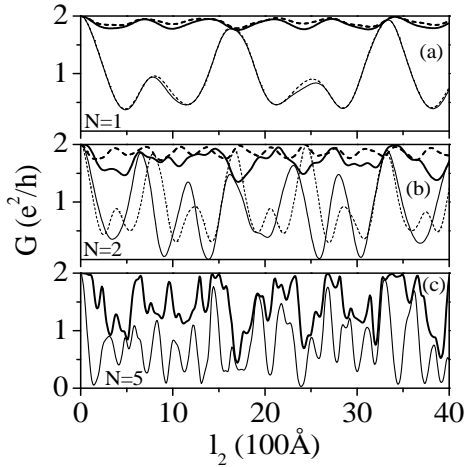}
\caption{The total (thick curves) and the spin-up (thin curves) conductance are illustrated as functions of
the SOI-segment's
length $l_2$ for electrons incident with energy $E=9$ meV. 
The solid curves in (a), (b) ,
and (c) are the results for a structure of 1,  2,
and  5 SOI segments, respectively, with $l_1=1250$\AA.
The dotted curves in (b) constitute the two-segment result with $l_1=1000$\AA.
The dotted curves in (a) is the double of the conductance obtained by assuming
electrons are
incident from only the first subband.
}
\label{fig5}
\end{figure}

In Fig. \ref{fig5} we calculate the total conductance $G$ (thick curves)
and the spin-up conductance $G^+$ (thin curves) as
functions of the SOI-segment length $l_2$ through periodic structures consisting of (a)
$N=1$, (b) $N=2$ and (c) $N=5$ units 
for a Fermi energy $E_F=E=9$ meV close to the bottom of the second subband, where strong
mismatch of the wave
functions of the second subband is expected at the interfaces.
For $N=1$ the transmission shows a periodic pattern and the reflection is
limited.
In contrast to the low-energy
transmission discussed in the previous subsection, the spin-up conductance here
does not %touch zero
vanish, i.e. the spin cannot be inverted completely by
the structure, because the spin of electrons in the first subband precesses
with a frequency twice that in the second subband at this energy.
Even if we assume that the electrons are incident only from the lowest subband,
i.e. by letting
$c_{2I}^{\pm}=0$ when carrying out the calculation, a similar result is obtained as shown by
the thick and thin dotted curves (doubled) in Fig. \ref{fig5}(a).
It is worthy of noting that
the transmission estimation by assuming one-mode incidence as
used in Ref.\onlinecite{egue}
is a good approximation for one-SOI-segment systems though not for multiple-SOI-segment
ones. In the above structure if we put 
two
identical SOI segments and separate
them by a segment without SOI of length $l_1$, a similar pattern of
the spin-up conductance as above appears but the oscillation
frequency is almost doubled as 
shown in Fig. \ref{fig5}(b), where the results corresponding to
$l_1=1250$\AA\ (solid curves) and $l_1=1000$\AA\ (dotted curves) are plotted.
The segment without SOi can change the relative phase of electrons between interfaces
and then the relative
spin orientation between electrons in the first and second subbands, therefore the initial
spin-up electrons can be totally spin-flipped by
the structure with $l_1=1250$\AA\ but not by that with $l_1=1000$\AA\ as
the thin curves indicate in Fig. \ref{fig5} (b).
With increasing number of SOI segments or 
SOI interfaces, the modulation of 
the conductance becomes complex and develops into conductance 
fluctuation. As an example, we plot the results for a five-SOI-segment structure in 
Fig. \ref{fig5}(c). The total conductance oscillates or fluctuates as a function of $l_2$
and the corresponding amplitude increases with increasing number of units. Strong
reflection happens when the number of the units is large enough and the multi-SOI-segment
structures can be used as electron filters even if two modes exist.

\subsection{The anticrossing-energy limit}

At the anticrossing energy $E_c$, opposite spin states from the first and
second
subbands mix strongly with each
other. Nevertheless, simple transmission patterns can be found even in
structures with multiple SOI segments.
In a quantum wire with one SOI segment of length $l_2=l$ sandwiched between
two segments (ends) without SOI,
the coefficient of
the output wave function in the anticrossing-energy limit is obtained
approximately with the help of Eqs. (\ref{cross}) and (\ref{tmatrix}):
\begin{equation}
\left( 
\begin{array}{c}
c_{1O}^{+} \\ 
c_{1O}^{-} \\ 
c_{2O}^{+} \\ 
c_{2O}^{-}
\end{array}
\right) =e^{ik_{c}l}\left( 
\begin{array}{c}
c_{1I}^{+}\cos (k_{\delta }l)-ic_{2I}^{-}\sqrt{k_2^0/k_1^0}\sin (k_{\delta
}l) \\ 
c_{1I}^{-}\exp (ik_{\alpha }l) \\ 
c_{2I}^{+}\exp (-ik_{\alpha }l) \\ 
-ic_{1I}^{+}\sqrt{k_2^0/k_1^0}\sin (k_{\delta }l)+c_{2I}^{-}\cos
(k_{\delta }l)
\end{array}
\right) 
=\frac{e^{ik_{c}l}}{\sqrt{2}}
\left( 
\begin{array}{c}
e^{-ik_{\delta }l}/\sqrt{k_1^0} \\ 
e^{ik_{\alpha }l}/\sqrt{k_1^0} \\ 
e^{-ik_{\alpha }l}/\sqrt{k_2^0} \\ 
e^{-ik_{\delta }l}/\sqrt{k_2^0}
\end{array}
\right).
\end{equation}
Here the normalized  coefficients $(c_{1I}^{\pm }=1/\sqrt{%
2k_1^0},c_{2I}^{\pm }=1/\sqrt{2k_2^0})$ of the incident waves
are used to get the rhs of the above equation.
The spin-up (spin-down) transmission then reads
\begin{equation}
T^{\pm}=1\pm \cos(k_{\alpha }l)\cos(k_{\delta }l)
\label{transm}
\end{equation}
with $T^{\pm}_1=1/2\pm 1/2\cos(k_{\alpha}+k_{\delta})l$
the transmission out of the first subband and 
$T^{\pm}_2=1/2\pm 1/2\cos(k_{\alpha}-k_{\delta})l$ out of the second subband.
For an rational ratio of
$k_{\alpha}/k_{\delta}\approx 9\pi^2\hbar^2/(16m^*\alpha w)$
the transmission is approximately a
periodic function of the length of the SOI segment.

\begin{figure}[h]
\includegraphics*[width=100mm,height=90mm]{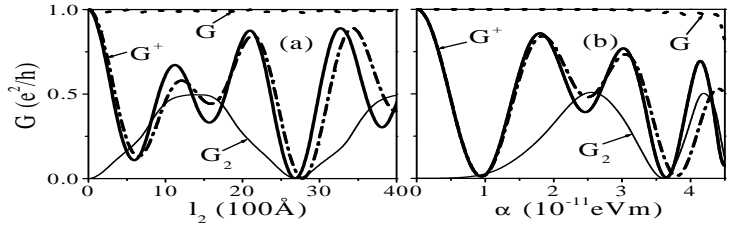}
\caption{The conductance
is plotted as a function of the length of the SOI segment (a) and of the SOI
strength $\alpha$ (b) when electrons of the anticrossing energy
of the SOI segment are assumed incident from only the first subband. The thick solid, thin solid, and dotted curves present the spin-up, second-subband, and total conductances respectively obtained numerically. As comparison, the dash-dotted curves show the spin-up conductance obtained from Eq. (\ref{transm}).
}
\label{fig6}
\end{figure}

Assuming electrons are incident only  from the first subband
($c_{2I}^{\pm }=0)$, we get the same approximate transmission as Eq. (\ref{transm}) but divided by two.
This result is also the same as that found
in Ref.  \onlinecite{egue} where the interface effect is neglected.
In Fig. \ref{fig6}(a)
we show the conductance versus  $l$ and in Fig. \ref{fig6}(b) versus $\alpha $. The Fermi energy of the
electron gas is equal to the anticrossing energy
$E_{c}(\alpha)$ and the electrons are assumed incident from only the
first subband. The dash-dotted and the thick solid curves represent the spin-up conductance
given respectively by the simplified expression Eq. (\ref{transm}) and by the
numerical solution of
Eqs. (\ref{eqsq}), (\ref{matrixs}), and (\ref{matrixq}).
When the anticrossing energy locates well between the minima of the second and the third subbands, corresponding to
$2.5<\alpha<3.5\times 10^{-11}$eVm here, the total transmission, shown by the dotted curves, is almost unity and
Eq. (\ref{transm}) can be used to estimate the spin transmission
through a quantum wire with a SOI segment of length up to several thousand angstroms.
Nevertheless, when $k_{\delta}l \sim n \pi/2$ for odd number $n$ almost half of the
electrons transit from the first subband to the second
subband (corresponding to $G_2 \sim 0.5 e^2/h$ shown by the thin solid curves
in Fig. \ref{fig6}) and Eq. (\ref{transm}) is in
poorer agreement with the numerical result. For a lower anticrossing energy or $\alpha>4\times 10^{-11}$eVm
the reflection due to SOI interface becomes important and Eq. (\ref{transm}) becomes less reliable. For
$\alpha<2.5\times 10^{-11}$eVm, the corresponding anticrossing energy is higher than the third
subband and the result given in Fig. \ref{fig6}(b) should be corrected by considering the effect of the third subband.

Actually, electrons should be incident simultaneously from both the
first and the second subband if the quantum wire is connected to
a Fermi electron reservoir with Fermi energy
higher then the second-subband bottom. At the anticrossing energy,
the wave functions of the first and the 
second subband mix equally with each other and electrons propagate almost equally through each subband.
The spin-up and total conductances
have quite similar dependence on $l_2$ and $\alpha$ as those plotted in Fig. \ref{fig6} with doubled values. In fact, even when the
Fermi energy is far away from the anticrossing energy but in the second subband,
the spin-up and spin-down
conductance $G^{\pm}$ of a Fermi gas can be well estimated by doubling the one
obtained by assuming incidence
from one subband as illustrated in Fig. \ref{fig5} (a) where the Fermi energy is close to
$\varepsilon_2(0)$.
Comparing the conductance
pattern in Fig. \ref{fig5} (a) with that in Fig. \ref{fig6} (a), we see mainly two
different features. At first, the reflection of electrons at $E_c$ is much lower because
the mismatch of wave function is less. Secondly, both the spin-up and the spin-down
conductances at $E_c$ can
vanish but only the spin-down one vanishes
at energy near $\varepsilon_2(0)$. This is a result of the fact that
the ratio between the precession
frequencies of electrons in the first and the second subbands changes
with the electron energy. The ratio is 5/3 in Fig. \ref{fig6} (a)
comparing to 2 in Fig. \ref{fig5} (a).

%fig7
\begin{figure}[h]
\includegraphics*[width=100mm,height=120mm]{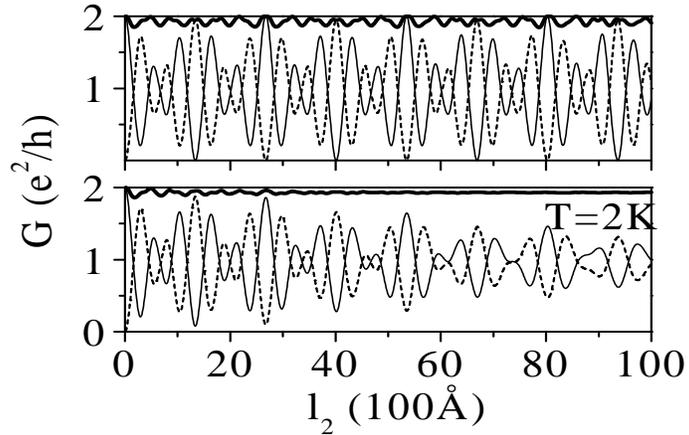}
\caption{Conductance through a two-SOI-segment quantum wire
as a function of $l_2$ at zero temperature (upper plot)
and at temperature $T=2K$ (lower plot). $G$ (thick solid curves),
$G^+$ (thin solid), and $G^-$ (dotted) are shown.
The parameters are chosen such that
$k_{\alpha}l_2=4k_{\delta}l_2=4k_0l_1=8\pi$
for $l_2=5352$\AA.
}
\label{fig7}
\end{figure}

In quantum wires with two SOI segments, the spin transmission at zero temperature is approximately
expressed as
\begin{equation}
T^{\pm}\approx 1\pm \cos k_0l_1 \cos (2k_{\delta}l_2) \cos (k_0l_1+2k_{\alpha}l_2)
\pm \sin k_0l_1 \sin(k_0l_1+2k_{\alpha}l_2).
\label{twoseg}
\end{equation}
Here $k_0=(k_1^0-k_2^0)/2$, with $k_1^0$ and $k_2^0$ the wave vectors of the first and second subband in the segment
without SOI.
The conductance as a function of $l_2$ shows a periodic pattern if the ratio
$k_{\alpha}/k_{\delta}$ is rational.
In Fig. \ref{fig7}, we show the conductance as a function of the
length of the SOI segment $l_2$ at zero temperature (upper panel) and at
temperature $T=2K$ (lower panel).
We choose the wire width as $w=609$\AA\ so that $k_{\alpha}=2k_0=4 k_{\delta}$, the electron
Fermi energy $E_F=E_c=11$meV, and the length of the segment without SOI
between the two SOI
segments $l_1=2676$\AA\ to satisfy $k_0l_1=2\pi$. In the zero temperature panel, we see a
periodic beating pattern similar to that of the one-segment wire plotted in
Fig. \ref{fig6} (a) but with a period $l_1=2676$\AA\ half of the one-segment one.
The origin of the beating
pattern is the difference of the spin-precession frequencies in different subbands: Because
$k_0l_1=2\pi$, Eq. (\ref{twoseg}) reduces to Eq. (\ref{transm}) and the partial conductance
$G_1^{\pm}\sim 1+\cos(10 k_{\delta}l_2)$ and
$G_2^{\pm}\sim 1+\cos(6 k_{\delta}l_2)$.
We note that the conductance pattern shows a sensitive
dependence on the length of segment without SOI.
If $l_1=2000$\AA is chosen, for example,
the numerical result shows that both $G_1^{\pm}$ and $G_2^{\pm}$ have a dependence close to
$1+\cos(8 k_{\delta}l_2)$ and the conductance pattern becomes completely different.
At finite temperature, as shown in the lower panel where $T=2$K,
the output polarization of
the electron current and the amplitude of the beating pattern decrease with
the length of the
structure as pointed out also in Ref. \onlinecite{mir}. The oscillation
amplitude becomes half of the
initial value at $l_2=8000$\AA\ so we can estimate that the depolarization length due to
the subband mixing in this structure is of the order of micrometer at $T=2$K.

\section{conclusion}

We have studied the SOI-interface effect on the ballistic spin transport through quantum
wires composed of a series of
segments with and without SOI. At low electron density, when the Fermi energy 
of the electron gas is much lower than the
second subband, the total conductance is modulated sinusoidally but 
the outgoing spin orientation remain the same as
that without SOI
interfaces. In periodic structures the modulation of the total transmission
develops into square gaps when
the length of the SOI segments
or the electron energy are varied. This feature is similar
to that obtained previously in stubbed waveguides \cite{wan,wan1}
with constant strength $\alpha$ everywhere
and has potential
applications in
establishing a spin transistor. 

At higher density, when two subbands are 
occupied, the outgoing spin orientation is further modulated due to the SOI-induced subband  mixing.
For electrons  with energy close to the anticrossing one
the transmission pattern
is approximately periodic as function of the length of the SOI segments
if the intrasubband 
SOI splitting $k_{\alpha}$ is a
rational multiple of the
intersubband one
$k_{\delta}$, though generally 
the two-subband transmission patterns are complex.
The reflection resulting from the SOI interfaces can be very significant when 
several SOI segments exist along the
quantum wire separated by non-SOI segments.
In this case
the transmission and  outgoing spin
orientation can be sensitive to the
length of the non-SOI segments as well as to that of the SOI segments.

Finally, the theoretical treatment presented here can also be used to study
quantum spin transport in quantum wires in
which
the SOI strength varies continuously.
To do so one simply has to divide the wire into a series of segments,
inside which the SOI strength can be treated approximately as constant,
with a different value from segment to segment.

\section{Acknowledgments}
The author thanks Prof. P. Vasilopoulos for stimulating discussions and useful
suggestions. This work was supported by the Canadian NSERC Grant No. OGP0121756.

\end{document}